\documentstyle[11pt,newpasp,twoside,epsf]{article}
\def\edcomment#1{\iffalse\marginpar{\raggedright\sl#1\/}\else\relax\fi}
\reversemarginpar

\begin{document}
\title{Dense Molecular Gas and Star Formation in Nearby Seyfert Galaxies}
\author{K. Kohno}
\affil{Institute of Astronomy, University of Tokyo, 
2-21-1, Osawa, Mitaka, Tokyo, 181-8588, Japan}
%  on, The Full Address of My Institution,
%Run in Address to Make Minimum Number of Lines}
\author{S. Matsushita}
\affil{Submillimeter Array, Harvard-Smithsonian Center for Astrophysics,
        P.O. Box 824, Hilo, HI 96721-0824, USA}
\author{B. Vila-Vilar\'o}
\affil{Steward Observatory, University of Arizona, Tucson, AZ 85721, USA}
\author{S. K. Okumura, T. Shibatsuka, M. Okiura}
\affil{Nobeyama Radio Observatory, Minamisaku, Nagano, 384-1305, Japan}
\author{S. Ishizuki, R. Kawabe}
\affil{National Astronomical Observatory, Mitaka, Tokyo 181-8588, Japan}

\begin{abstract}
An imaging survey of CO(1$-$0), HCN(1$-$0), 
and HCO$^+$(1$-$0) lines in the centers of nearby Seyfert galaxies 
has been conducted
using the Nobeyama Millimeter Array and the RAINBOW interferometer.
Preliminary results reveal that 3 Seyferts out of 7 show abnormally high HCN/CO 
and HCN/HCO$^+$ ratios, which cannot occur even in nuclear starburst
galaxies. We suggest that the enhanced HCN emission 
originated from X-ray irradiated dense obscuring tori,
and that these molecular line ratios can be a new diagnostic
tool to search for ``pure'' AGNs. According to our HCN diagram,
we suggest that NGC 1068, NGC 1097, and NGC 5194 host
``pure'' AGNs, whereas Seyfert nuclei of
NGC 3079, NGC 6764, and NGC 7469 may be ``composite'' in nature. 
\end{abstract}

\section{Introduction}

Dense molecular matter is considered to play various roles
in the vicinity of active galactic nuclei (AGNs).
The presence of dense and dusty interstellar matter (ISM),
which obscures the broad line regions in AGNs,
is inevitable at a few pc - a few 10 pc scale
according to the unified model of Seyfert galaxies.
This circumnuclear dense ISM
could be a reservoir of fuel for nuclear activity,
and also be a site of massive star formation.
%(e.g. Cid Fernandes et al.\ 2001).

In order to investigate
dense molecular matter in the centers of Seyfert galaxies,
we have conducted an imaging survey of CO(1$-$0), HCN(1$-$0),
and HCO$^+$(1$-$0) lines in nearby Seyfert galaxies
using the Nobeyama Millimeter Array (NMA).
High resolution HCN observations of Seyfert galaxies
are of interest because unusually strong HCN emission
has been reported in the type-2 Seyfert galaxies NGC 1068
(Jackson et al. 1993; Tacconi et al. 1994; Helfer \& Blitz 1995)
and NGC 5194 (Kohno et al.\ 1996).
The HCN/CO integrated
intensity ratios in brightness temperature scale,
$R_{\rm HCN/CO}$ hereafter, 
within the central $r \sim$ a few 10 pc 
region exceed 0.4, which {\it is never observed in non-Seyfert
galaxies} including nuclear starburst galaxies ($R_{\rm HCN/CO}
< 0.3$; see Kohno et al.\ 1999 and references therein).

%In the following sections, we show preliminary maps of 
%CO, HCN, and HCO$^+$ lines, including a detailed report of
%NGC 1097 where strong enhancement of HCN was detected.
%Interpretation and implication are discussed in Section 4.

\section{CO, HCN, and HCO$^+$ Images of Seyferts}

Figure 1 shows a part of the CO images of Seyfert galaxies
obtained with the NMA so far.
Some Seyfert galaxies have also been observed
using the RAINBOW interferometer, which is a 7 elements 
combined array consisting of six 10 m dishes (NMA) and NRO
45 m telescope; see Sofue et al.\ (2001) for NGC 3079,
and see Okiura et al.\ (2001, this volume) for NGC 7469.

These CO images show a wide variety of gas morphologies
in the central kpc regions of Seyferts, just as in the case
of {\it normal} spirals (Sakamoto et al.\ 1999).
Single dish CO surveys aiming to determine the
global amount of molecular gas in Seyferts have
revealed that there is no significant difference
between the {\it total} amounts of molecular gas in
Seyfert and quiescent spirals (e.g., Vila-Vilar\'o et al. 1998).
Our higher angular resolution CO images
may already suggest that the accumulation of molecular gas
in the central kpc region is still insufficient for Seyfert
activity.

\begin{figure}
\plotone{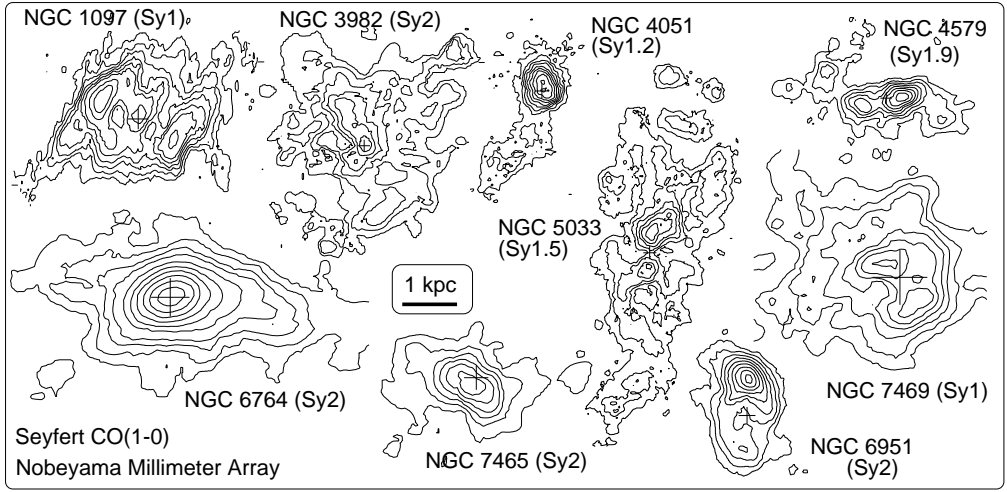}
\caption{
CO(1$-$0) distributions in the central kpc regions of
nearby Seyfert galaxies, displayed with the same linear scale.  
}
\end{figure}

\begin{figure}
\plotone{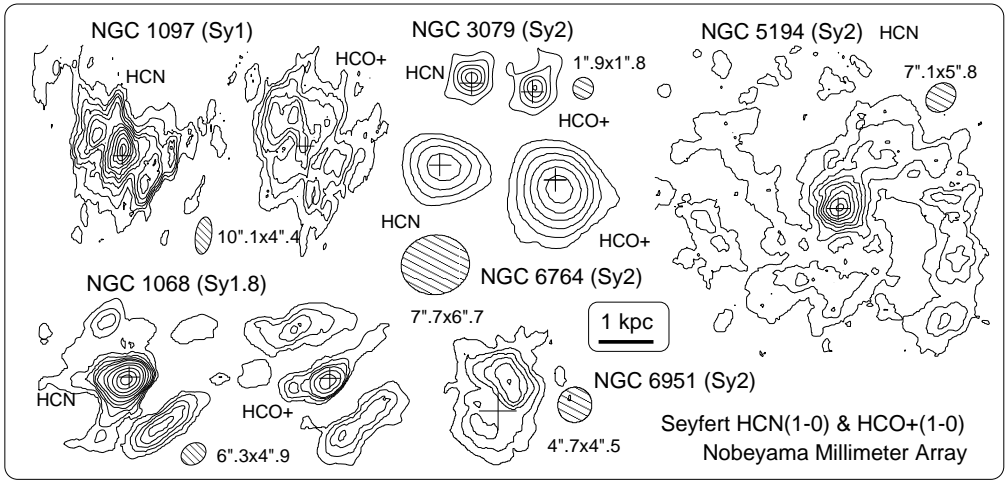}
\caption{
HCN(1$-$0) and HCO$^+$ distributions in the central kpc regions of
nearby Seyfert galaxies with synthesized beams indicated.
}
\end{figure}

%\section{HCN and HCO$^+$ Images}

\begin{figure}
\plotone{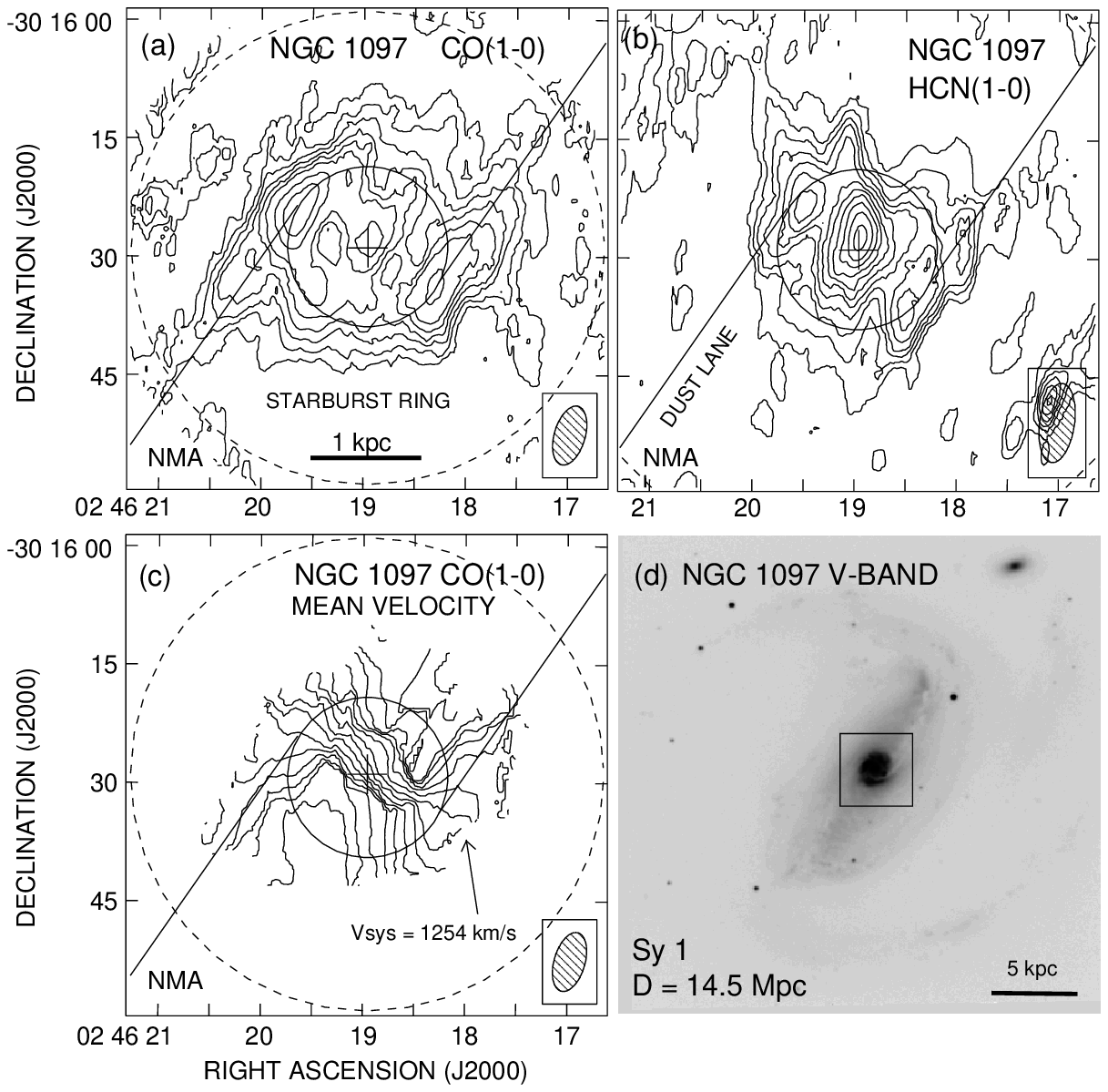}
\caption{
The CO(1$-$0) and HCN(1$-$0) images
of the type-1 Seyfert NGC 1097 taken with the NMA.
The central cross in (a), (b), and (c) marks
the position of the nucleus (6 cm continuum peak; Hummel et al.\ 1987).
%In each molecular map, the NMA field of view
%($60''$ for CO and $80''$ for HCN)
%is indicated by a dashed circle.
The circumnuclear starburst ring
with a radius of $r \sim 10''$ or 700 pc, clearly traced with
H$\alpha$ and radio continuum (e.g. Hummel et al.\ 1987)
and dust lanes along the bar (see (d)) are indicated.
(a) CO(1$-$0) integrated intensity map.
The synthesized beam is
$7.\hspace{-2pt}''7 \times 3.\hspace{-2pt}''9$
($540 \times 270$ pc) with a P.A. of $-16^{\circ}$.
Contour levels are 1.5, 3, 4.5, 6, 7.5, 9, 12, 15, and 18 $\sigma$,
where 1 $\sigma$ = 3.39 Jy beam$^{-1}$ km s$^{-1}$
or 10.4 K km s$^{-1}$ in $T_{\rm b}$. 
This corresponds to a face-on
gas surface density $\Sigma_{\rm gas}$ of 34.7 $M_\odot$ pc$^{-2}$,
calculated as $\Sigma_{\rm gas} = 1.36 \times \Sigma_{\rm H_2}$
and $\Sigma_{\rm H_2} =
4.81 \times (I_{\rm CO}/\mbox{K km s$^{-1}$}) \cdot cos(i) \cdot
(X_{\rm CO}/(3.0 \times 10^{20} \mbox{cm$^{-2}$ (K km s$^{-1}$)$^{-1}$})$,
where $i = 46^\circ$.
%where $X_{\rm CO}$ is the CO - to N(H$_2$) conversion factor.
%(e.g. Scoville et al. 1987; Solomon et al. 1987),
%and $I_{\rm CO}$ is the velocity integrated CO intensity,
%and $i$ is the inclination of the disk ($46^\circ$ for NGC 1097;
%Ondrechen et al.\ 1989).
The ``CO twin peaks'' is evident, as well as the nuclear CO source.
%Attenuation due to primary beam pattern of each 10 m dish
%has been corrected in this map.
%
(b) HCN(1$-$0) integrated intensity.
The synthesized beam is
$10.\hspace{-2pt}''1 \times 4.\hspace{-2pt}''4$
($710 \times 310$ pc) with a P.A. of $-8^{\circ}$.
Contour levels are 1.5, 3, 4.5, $\cdots$, and 16.5 $\sigma$,
where 1 $\sigma$ = 0.711 Jy beam$^{-1}$ km s$^{-1}$
or 3.54 K km s$^{-1}$ in $T_{\rm b}$. 
Most of the HCN
emission is dominated by the central unresolved peak.
%Attenuation due to primary beam pattern of each 10 m dish
%has been corrected in this map.
%
(c) Intensity-weighted mean velocity map of CO(1$-$0).
The contour interval is 30 km s$^{-1}$.
%, and the systemic velocity of
%1524 km s$^{-1}$, determined by the fitting of this data,
%is indicated.
%Circular rotation seems to dominate the kinematics in the central
%$r<10''$ region (i.e. within the circumnuclear ring) in NGC 1097,
%whereas very strong non-circular motion
Very strong non-circular motion along the bar
is suggested by isovelocity contours parallel to the dust lanes.
(d) V-band image of $400'' \times 400''$ (28 kpc $\times$ 28 kpc)
region of NGC 1097 (Quillen et al.\ 1995).
%A solid box shows the mapped region with the NMA ($60'' \times 60''$).
%Intensity scale ($\gamma$ factor) of this image
%has been modified to enhance fine structures.
%
}
\end{figure}

Figure 2 shows the HCN and HCO$^+$ images of Seyfert galaxies.
Except for NGC 5194 and NGC 6951,
HCN and HCO$^+$ lines were observed simultaneously,
thanks to the wide (1024 MHz) band width of a new spectro-correlator
(UWBC; Okumura et al.\ 2000).
This enables us to measure HCN/HCO$^+$ ratios 
($R_{\rm HCN/HCO^+}$) accurately
(systematic error must be less than a few \%).

We find significant enhancement of HCN
toward the nucleus of the type-1 Seyfert galaxy, NGC 1097
(Storchi-Bergmann et al.\ 1997). As demonstrated in Figure 3 and 4, 
the $R_{\rm HCN/CO}$ in the center of NGC 1097 is enhanced
up to 0.34. This is the 3rd detection of abnormally 
($R_{\rm HCN/CO} > 0.3$) enhanced HCN
after NGC 1068 and NGC 5194.
We have already mapped another 4 Seyferts (NGC 3079, NGC 6764, 
NGC 6951, and NGC 7469). We therefore find that 3 Seyferts out of 7 show
extreme enhancement of HCN based on our preliminary data.

Another intriguing point is the remarkable weakness of HCO$^+$ emission 
in NGC 1068 and NGC 1097; in Figure 2, we find that $R_{\rm HCN/HCO^+}$
is 2.3 and 2.1 in NGC 1068 and NGC 1097, respectively.
The ratios decrease to about unity in the circumnuclear
starburst region. Preliminary results of HCO$^+$ in M51
also show very weak HCO$^+$ 
($R_{\rm HCN/HCO^+} > 2$; Shibatsuka et al., in preparation).

\section{Discussions}

\begin{figure}
\plotone{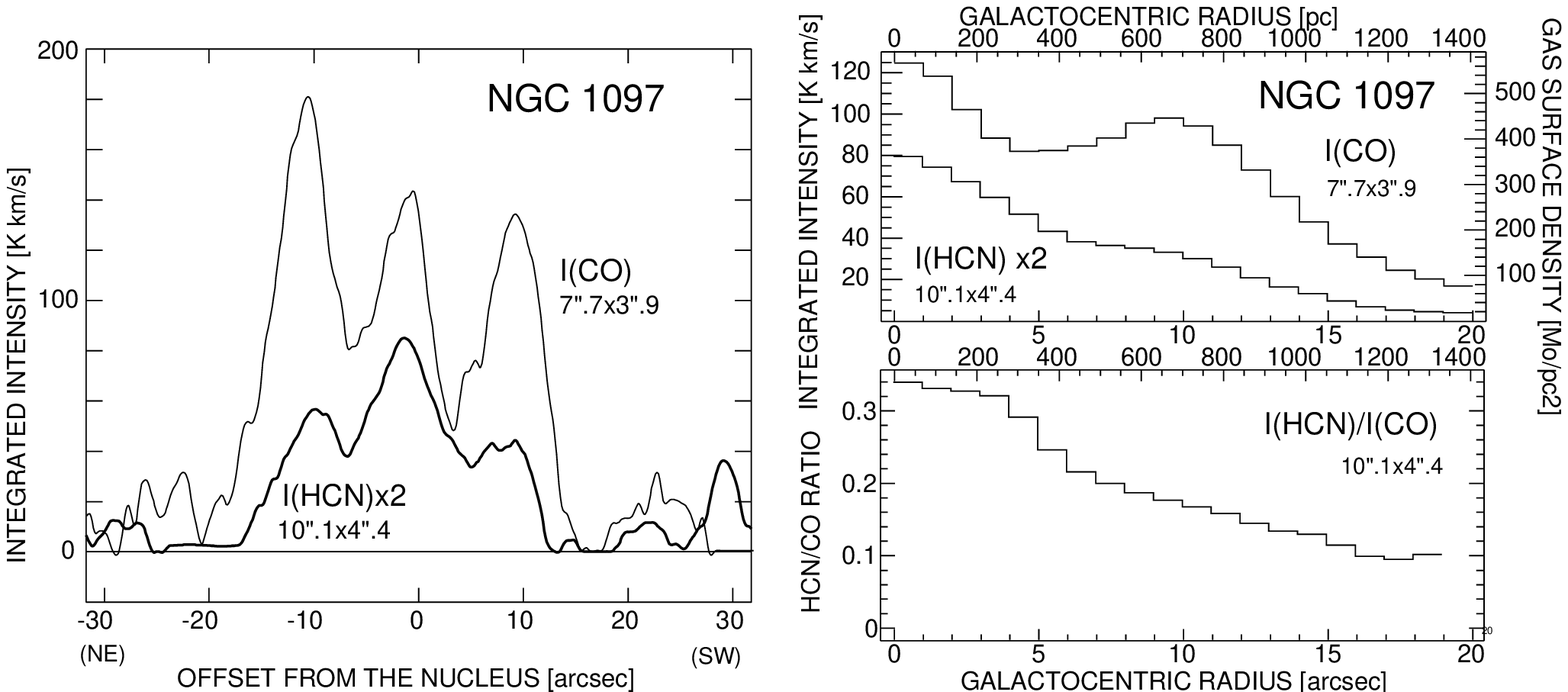}
\caption{
CO and HCN intensity profiles through the nucleus
along P.A. = 60$^\circ$ line (left), azimuthally
averaged radial distributions of CO and HCN (top right),
and HCN/CO ratio (bottom right).
}
\end{figure}

We have observed extremely strong HCN emission
in 3 Seyferts out of 7.
What is the nature of these ``HCN-enhanced Seyferts''?
Here we compare the observed line ratios in Seyferts
with those in nuclear starburst galaxies,
which were also measured with similar angular resolutions.
In Figure 5, it is immediately evident that Seyferts without abnormal
HCN enhancements,  
i.e. NGC 3079, NGC 6764, and NGC 7469, show
$R_{\rm HCN/CO}$ and $R_{\rm HCN/HCO^+}$ values
just comparable to those in nuclear starbursts; they have 
$R_{\rm HCN/CO}$ less than 0.3, and $R_{\rm HCN/HCO^+}$
ranging from 0.5 to 1.5.
On the other hand, HCN-enhanced Seyferts, i.e.
NGC 1068 and NGC 1097, also have very high $R_{\rm HCN/HCO^+}$
values ($> 2$). Note that Nguyen-Q-Rieu et al. (1992)
reported a very high $R_{\rm HCN/HCO^+}$ in NGC 3079 
and Maffei 2 ($>3$), yet our new simultaneous measurements gave
moderate ($\sim 1$) ratios.

We propose that these two groups in our ``HCN diagram''
(Figure 5) can be understood 
in terms of ``AGN - nuclear starburst connection''
(note that this should not be confused with ``AGN - starburst
cohabitation'', which often refers to the association
of AGN with star formation on galactic scales in AGN hosts).
In the Seyferts with line ratios comparable to those
in nuclear starburst galaxies, it seems likely
that nuclear starburst (presumably in the dense molecular torus)
is associated with the Seyfert
nucleus (i.e., ``composite'').
In the nuclear regions of composite Seyferts,
HCO$^+$ fractional abundance is expected to 
increase due to frequent supernova (SN) explosions.
In fact, in evolved starbursts such as
M82, where large scale outflows
have occurred due to numerous
SN explosions, 
HCO$^+$ is often stronger than HCN
(e.g. Nguyen-Q-Rieu et al.\ 1992). 
On the other hand, 
the HCN-enhanced Seyferts, which shows
$R_{\rm HCN/CO} > 0.3$ and $R_{\rm HCN/HCO^+} > 2$,
would host ``pure'' AGNs, 
where there is no associated nuclear starburst activity. 
In such a condition, the HCN line can be very strong because
it has been predicted that fractional abundance of
HCN is enhanced by strong X-ray radiation
from AGN (Leep \& Dalgarno 1996), resulting 
in abnormally high $R_{\rm HCN/CO}$ and $R_{\rm HCN/HCO^+}$
values.
Our interpretation is supported
by other wavelength data; for instance,
NGC 1068 has been claimed as a pure Seyfert (Cid Fernandes et al.\ 2001
and references therein), whereas NGC 6764 (Schinnerer et al.\ 2000)
and NGC 7469 (Genzel et al.\ 1995) have a composite nature.
We need further analysis to validate
the proposed interpretation, but if it is the case,
this will serve as a new way to investigate the nature
of AGNs; 
although this technique requires high angular resolution observations
in order to avoid contaminations from extended 
circumnuclear star-forming regions, it has some
advantages (e.g., not being affected by dust extinction).

%In summary, 
%we suggest that the abnormally enhanced HCN emission would be 
%originated from X-ray irradiated dense obscuring torus,
%and these molecular line ratios can be a new diagnostic
%tool to search for ``pure'' AGN. According to our HCN diagram,
%we suggest that NGC 1068, NGC 1097, and NGC 5194 would host
%a ``pure'' AGN, whereas Seyfert nuclei of
%NGC 3079, NGC 6764, and NGC 7469 may be
%associated with nuclear starbursts (i.e. ``composite'').

\begin{figure}
\plotone{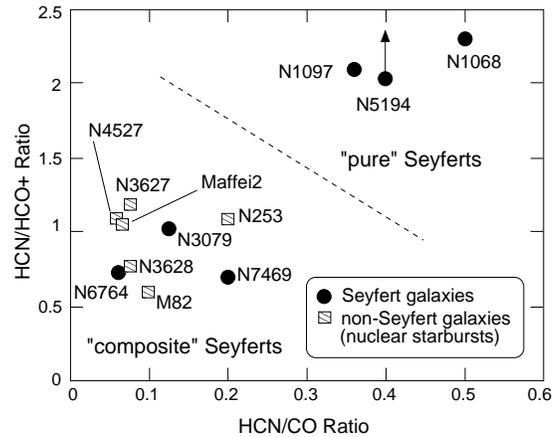}
%\plotfiddle{m82-laPalma-fig1.eps}{4.0cm}{0}{45}{45}{-200}{-20}
\caption{
Molecular line ratios in Seyfert and non-Seyfert (nuclear starburst) galaxies.
%Some Seyferts show $R_{\rm HCN/CO}$ and $R_{\rm HCN/HCO^+}$ values
%comparable to those in nuclear starburst galaxies, whereas
%both $R_{\rm HCN/CO}$ and $R_{\rm HCN/HCO^+}$ in
%NGC 1068, NGC 1097, and NGC 5194 are extremely high.
%Note that NGC 3627 may have a Seyfert nucleus; classified as T2/S2 
%by Ho et al. (1997).
}
\end{figure}

\acknowledgements
We are indebted to the NRO staff for their efforts in improving
the performance of the array.

\end{document}